# Phonon renormalization induced by electric field in ferroelectric P(VDF-TrFE) nanofibers

Lan Dong,[1] Qing Xi,[1] Jun Zhou,[1]* Xiangfan Xu,[1]* and Baowen Li[2]

[1]Center for Phononics and Thermal Energy Science, China-EU Joint Center for Nanophononics, School of Physics Science and Engineering, Tongji University, Shanghai 200092, China

[2]Department of Mechanical Engineering and Department of Physics, University of Colorado at Boulder, CO 80309, USA

**ABSTRACT:** We report phonon renormalization induced by an external electric field $E$ in ferroelectric poly(vinylidene fluoride-trifluoroethylene) [P(VDF-TrFE)] nanofibers through measuring the $E$-dependent thermal conductivity. Our experimental results are in excellent agreement with the theoretical ones derived from the lattice dynamics. The renormalization is attributed to the anharmonicity that modifies the phonon spectrum when the atoms are pulled away from their equilibrium positions by the electric field. Our finding provides an efficient way to manipulate the thermal conductivity by tuning external fields in ferroelectric materials.

The conventional phonon theory, that considers the anharmonicity as a perturbation, explains very well the behavior of thermal conductivity in crystals at low temperature. However, when the material is subject to an intensive external electric/magnetic field, or stress, or high temperature, the atoms deviate far away from their original equilibrium positions and consequently induce a strong anharmonicity that significantly modifies the phonon spectrum, and the perturbation theory is no longer valid. For example, at high temperature, thermal conductivities of some crystals are not inversely proportional to temperature, which is mainly attributed to large thermal motion of atoms at high temperature [1, 2].

To incorporate the effect of anharmonicity to the conventional phonon transport theory, people usually renormalize the vibrational modes [3, 4]. These renormalized vibrational modes are called *phonon renormalization*, which is more obvious in materials containing 'rattlers', as those 'rattlers' have much larger displacements than other atoms [5]. Another example to illustrate phonon renormalization is exposing the materials to a stress, since stress could directly change the lattice structure of crystal and thus alter the elastic constants [6, 7], that lead to the change of thermal conductivity. Indeed, tuning thermal conductivity via stress has been widely adopted due to its general applicability to materials, ranging from nanostructures [8-11], insulating solids [12], silicon nanowires [13], semiconductor nanofilms [14], to organic polymers [7, 15]. However, manipulating the thermal conductivity through changing temperature and stress are not easy for practical applications. People turn to other options like applying electric and/or magnetic fields [16, 17]. However, as far as we know that there is no systematic study on the phonon renormalization induced by external electric/magnetic field, neither from theoretical approach nor from experimental approach. Possible reason might be that the change of thermal conductivity due to phonon renormalization is too small to be measured in inorganic crystals.

Compared with inorganic materials, the change of thermal conductivity of organic materials could be easier to be observed since the molecule chains of polymers are flexible and bendable, and their thermal conductivity could be more sensitive to external field [15]. In case of ferroelectric polymers with large polarizability, the dipole motion driven by electric field is very likely to induce phonon renormalization. Among ferroelectric polymers, poly(vinylidene fluoride) (PVDF) and its copolymers are good prototype for their outstanding piezoelectric[18] and ferroelectric properties [19, 20].

In this Letter, we report, to the best of our knowledge, the first experimental observation of a tunable phonon renormalization in poly(vinylidene fluoride-trifluoroethylene) [P(VDF-TrFE)] nanofibers by measuring the electric field dependent thermal conductivity. The samples are fabricated by the electrospinning method and the thermal conductivity is then measured by the thermal bridge method. We find that the measured electric field dependence of thermal conductivity, due to phonon renormalization, matches well with the analytically derived formula which considers the piezoelectric effect and the anharmonicity corresponding to thermal expansion. This finding provides a new way to manipulate the thermal conductivity in ferroelectric materials.

Figure 1 illustrates the concept of phonon renormalization induced by electric field. As an example, one-dimensional (1D) diatomic chain with one positive charge and one negative charge in each unit cell is shown in Fig. 1(a). The lattice constant is $a$ and the separation between two atoms in each unit cell is $b$. The Taylor expansion of the potential energy can be written in

$$\Phi = \Phi^{(0)} + \frac{1}{2}\sum_{iv,i'v'}\Phi^{(2)}_{iv,i'v'}(X_{iv}-X^0_{iv})(X_{i'v'}-X^0_{i'v'})$$
$$+\frac{1}{6}\sum_{iv,i'v',i''v''}\Phi^{(3)}_{iv,i'v',i''v''}(X_{iv}-X^0_{iv})(X_{i'v'}-X^0_{i'v'})(X_{i''v''}-X^0_{i''v''})+\cdots \quad (1)$$

where $X_{iv} = R_i + d_v$ denotes the coordinate of the atom $(i,v)$ referring to the $v^{\text{th}}$ atom in $i^{\text{th}}$ unit cell, $R_i$ is the position vector of the $i^{\text{th}}$ unit cell and $d_v$ is the location of the $v^{\text{th}}$ atom with respect to $R_i$. $\Phi^{(0)}$ is a constant, $\Phi^{(n)}$ is the $n^{\text{th}}$ order derivative of the potential energy when atoms are located at their equilibrium positions $X^0_{iv}$. The potential of the positive atom near its equilibrium position is denoted by the solid curve in the lower graph of Fig. 1(a), and the harmonic term is specified in dashed curve. The force constants are determined by the harmonic term. It is convenient to qualitatively describe the quasi-1D molecular chain of P(VDF-TrFE) with 1D diatomic chain model by treating $CH_2$ and $CF_2$ clusters as positive charge and negative charge as shown in Fig. 1(c), respectively.

When an electric field is adiabatically applied to the 1D diatomic chain as shown in Fig. 1(b), atoms are pulled away from their original equilibrium positions to new ones, $\tilde{X}^0_{iv}$. The positive charge and the negative charge in each unit cell move oppositely. Consequently, the lattice constant changes from $a$ to $a'$ and the separation between two atoms in each unit cell changes from $b$ to $b'$. The renormalized potential energy with respect to the new equilibrium positions is

$$\tilde{\Phi} = \tilde{\Phi}^{(0)} + \frac{1}{2}\sum_{iv,i'v'}\tilde{\Phi}^{(2)}_{iv,i'v'}(X_{iv}-\tilde{X}^0_{iv})(X_{i'v'}-\tilde{X}^0_{i'v'})+\cdots, \quad (2)$$

where $\tilde{\Phi}^{(2)}_{iv,i'v'} = \left(\frac{\partial^2\tilde{\Phi}}{\partial X_{iv}\partial X_{i'v'}}\right)\Big|_0$ denotes the modified force constants. They can be calculated by comparing Eq. (1) and Eq. (2), and the nearest neighboring terms are listed as follows:

$$\tilde{\Phi}^{(2)}_{i+,i+} = \Phi^{(2)}_{i+,i+} + \frac{1}{2}\left[\Phi^{(3)}_{i+,i+,i+} - \Phi^{(3)}_{i+,i+,(i+1)-} - \Phi^{(3)}_{i-,i+,i+}\right](b'-b) + \Phi^{(3)}_{i+,i+,(i+1)-}(a'-a), (3a)$$

$$\widetilde{\Phi}^{(2)}_{i-,i+} = \Phi^{(2)}_{i-,i+} + \frac{1}{2}\left[\Phi^{(3)}_{(i-1)+,i-,i+} + \Phi^{(3)}_{i-,i+,i+} - \Phi^{(3)}_{i-,i+,(i+1)-} - \Phi^{(3)}_{i-,i-,i+}\right](b'-b)$$
$$+ \left[\Phi^{(3)}_{i-,i+,(i+1)-} - \Phi^{(3)}_{(i-1)+,i-,i+}\right](a'-a), \tag{3b}$$

$$\widetilde{\Phi}^{(2)}_{i+,(i+1)-} = \Phi^{(2)}_{i+,(i+1)-}$$
$$+ \frac{1}{2}\left[\Phi^{(3)}_{i+,i+,(i+1)-} + \Phi^{(3)}_{i+,(i+1)-,(i+1)+} - \Phi^{(3)}_{i+,(i+1)-,(i+1)-} - \Phi^{(3)}_{i-,i+,(i+1)-}\right]$$
$$\times (b'-b) + \left[\Phi^{(3)}_{i+,(i+1)-,(i+1)-} + \Phi^{(3)}_{i+,(i+1)-,(i+1)+}\right](a'-a), \tag{3c}$$

$$\widetilde{\Phi}^{(2)}_{i-,i-} = \Phi^{(2)}_{i-,i-} + \frac{1}{2}\left[-\Phi^{(3)}_{i-,i-,i-} + \Phi^{(3)}_{(i-1)+,i-,i-} + \Phi^{(3)}_{i-,i-,i+}\right](b'-b)$$
$$- \Phi^{(3)}_{(i-1)+,i-,i-}(a'-a), \tag{3d}$$

Considering only the nearest neighboring terms of the third order derivatives, the force constant in Eq. (3) can be rewritten as:

$$\widetilde{\Phi}^{(2)}_{i-,i+} = -\beta_1 + \delta_1(b'-b), \tag{4a}$$
$$\widetilde{\Phi}^{(2)}_{i+,(i+1)-} = -\beta_2 - \delta_2(b'-b) + \delta_2(a'-a), \tag{4b}$$
$$\widetilde{\Phi}^{(2)}_{i+,i+} = \widetilde{\Phi}^{(2)}_{i-,i-} = \beta_1 + \beta_2 + (\delta_2 - \delta_1)(b'-b) - \delta_2(a'-a). \tag{4c}$$

Here we denote $\Phi^{(2)}_{i+,i-} = -\beta_1$, $\Phi^{(2)}_{i+,(i+1)-} = -\beta_2$, $\Phi^{(2)}_{i+,i+} = \Phi^{(2)}_{i-,i-} = \beta_1 + \beta_2$, $\Phi^{(3)}_{i+,i+,i-} = -\Phi^{(3)}_{i+,i-,i-} = \delta_1$, $\Phi^{(3)}_{i+,i+,(i+1)-} = -\Phi^{(3)}_{i+,(i+1)-,(i+1)-} = -\delta_2$, $\Phi^{(3)}_{i+,i+,i+} = -\Phi^{(3)}_{i-,i-,i-} = -\delta_1 + \delta_2$. The relations $\sum_{i'v'} \Phi^{(2)}_{iv,i'v'} = 0$ and $\sum_{i'v',i''v''} \Phi^{(3)}_{iv,i'v',i''v''} = 0$ are used in the calculations.

The variation of force constants shown in Eq. (4) will change the sound velocity from $v_s = a\sqrt{\frac{\beta_1\beta_2}{(\beta_1+\beta_2)(M_++M_-)}}$ to

$$v'_s \approx v_s \frac{a'}{a}\sqrt{1 - \frac{\beta_2^2\delta_1 - \beta_1^2\delta_2}{2\beta_1\beta_2(\beta_1+\beta_2)}(b'-b) - \frac{\beta_1\delta_2}{\beta_2(\beta_1+\beta_2)}(a'-a)}, \tag{5}$$

with accuracy to the first order of the variation of $\Phi^{(2)}_{iv,i'v'}$. The variation of $a$ and $b$ induced by electric field could be estimated from piezoelectric effect and dielectric properties [21]

$$\frac{a'-a}{a} = d_{33}E, \quad b'-b = -\left(\frac{\Omega}{\overline{M}}\right)^{\frac{1}{2}} \frac{b_{12}}{b_{11}} E, \tag{6}$$

where $d_{33}$ is piezoelectric coefficient, $b_{11} = -\omega_0^2$, and $b_{12} = [\varepsilon(0) - \varepsilon(\infty)]^{1/2}\varepsilon_0^{1/2}\omega_0$. $\Omega$ is the volume per unit cell, $\overline{M} = M_+M_-/(M_++M_-)$ is the reduced mass, $\omega_0$ is the frequency of transverse optic mode, $\varepsilon_0$ is permittivity of free space, $\varepsilon(0)$ is the static dielectric constant, and $\varepsilon(\infty)$ the high-frequency dielectric constant. It is obvious that both $a'-a$ and $b'-b$ are linearly dependent on $E$. One can easily find that $v'_s \approx v_s(1+d_{33}E)\sqrt{1+\gamma E}$ where $\gamma$ can be determined from Eq. (5). A simplified symmetric case of $\Delta = \frac{\delta_1}{\beta_1^2} - \frac{\delta_2}{\beta_2^2} = 0$ leads to a

limiting form of $\gamma$ which shows

$$\gamma_s = \lim_{\Delta \to 0} \gamma = -\frac{v_s^2}{k_B}(M_- + M_+)\alpha_1 d_{33}, \tag{7}$$

where $\alpha_1 = \frac{k_B}{2a}\left(\frac{\delta_1}{\beta_1^2} + \frac{\delta_2}{\beta_2^2}\right)$ is the thermal expansion coefficient, $k_B$ is the Boltzmann coefficient. If we ignore the changes of scattering strength and specific heat, the thermal conductivity of polymer is proportional to its sound velocity [7]. Then the electric field dependence of thermal conductivity due to phonon renormalization is $\kappa \propto (1 + d_{33}E)\sqrt{1 + \gamma E}$. We point out that $\gamma$ should be in the same order of $\gamma_s$. The difference between them is attributed to the detailed molecular chain conformation in real polymer as illustrated in Fig. 1(d). The reason is that there are many other effects which have not been considered in our model: 1) bond angle variation; 2) chain orientation; 3) crystallinity; 4) phase transition of the polymer (there are two phases of P(VDF-TrFE)); 5) bond length variation [17, 22]. We notice that $\frac{v_s^2}{k_B}(M_- + M_+)\alpha_1 \gg 1$ is usually satisfied in ferroelectric polymers, then the thermal conductivity of ferroelectric polymer under electric field is

$$\kappa \approx \kappa_0\sqrt{1 + \gamma E}, \tag{8}$$

where $\kappa_0$ is the thermal conductivity in the absence of electric field.

We testify our phonon renormalization theory by studying the electric field dependence of thermal conductivity of P(VDF-TrFE) nanofibers. $\gamma_s$ of this material is estimated to be $3.5 \text{ nmV}^{-1}$ by using $d_{33} = -55 \times 10^{-3} \text{ nmV}^{-1}$ [23], $\alpha_1 = 6.6 \times 10^{-4} \text{ K}^{-1}$ [23], $M_- + M_+ = 2.31 \times 10^{-25}$ kg [24] and $v_s = 2400 \text{ ms}^{-1}$ [25]. Conventional PVDF is mainly composed by the non-polar gauche isomers (α-phase) which leads to week ferroelectricity [26]. The ferroelectricity of P(VDF-TrFE) can be enhanced by introducing more β-phase from larger polymerization ratio of trifluoroethylene (TrFE). In this experiment, we prepared P(VDF-TrFE) nanofibers of 70/30 molar ratios and implemented definite post-treatments. Fig. 2a exhibits XRD spectra of P(VDF-TrFE) powders and electrospinning nanofibers. The presence of a dominant peak at 2θ=20°, which corresponds to the β-phase [26], confirms that both powders and nanofibers possess intrinsic ferroelectricity. Fig. 2b exhibits that a suspended P(VDF-TrFE) nanofiber formed across the two suspended SiN$_x$ membranes by electrospinning technology. After electrospinning, the suspended P(VDF-TrFE) nanofibers need to be annealed at 140°C in N$_2$ circumstance to improve its crystallization [25]. The two SiN$_x$ membranes were covered by platinum (Pt) coils, acting as heater and temperature sensor for thermal conductivity measurement. Two Pt/SiN$_x$ electrodes at the middle of the whole micro-device was used to apply electric field along axis of nanofibers. To avoid the interaction between a plurality of nanofibers, we chose a suspended single nanofiber, which however is extremely difficult for electrospinning to measure the polarization tunable thermal conductivity. In order to observe the change of thermal conductivity caused by the external field, the whole suspended micro-device was placed in a cryostat with high vacuum on the order of $1\times10^{-5}$ Pa to reduce the thermal convection. Since the change of thermal conductivity of P(VDF-TrFE) nanofibers would be much lower than the measurement sensitivity of traditional thermal bridge method [27, 28], we adopted the differential circuit configuration in our experiments due to its advanced measurement sensitivity approach to 10 pWK$^{-1}$. [29, 30]

Two samples are prepared where the diameter and length of Sample 1 (Sample 2) are 138 nm (511nm) and 1.95 μm (1μm), respectively. The ferroelectricity of polymer is sensitive to its post-processing condition [31]. Therefore, we further treat Sample 1 with three different post-processing methods (Referred to as Sample 1a, Sample 1b and Sample 1c) before measurement. Sample 1a and Sample 2 are annealed in nitrogen at 140°C and followed by furnace cooling; Sample 1b is baked in high vacuum cryostat at 67 °C; Sample 1c is annealed in nitrogen at 140°C and followed by a slow cooling (2 hours), which is a recrystallization process.

The measured thermal conductivity of all the samples under positive and negative electric field at $T$ = 300K are shown in Fig. 3. It is shown that thermal conductivity increases with the electric field. The electric field dependence of measured thermal conductivity of Samples 1a, 1c and 2 are in good agreement with Eq. (8). The fitted values of $\gamma$ are listed in Table.1. For Sample 1a, $\gamma$ equals to 1.1 $\gamma_s$ for both positive and negative fields. The fitted $\gamma$ of Sample 1c, which is cooled slowly after annealing, is 2.8 $\gamma_s$ for positive field and 3.1 $\gamma_s$ for negative field. The larger $\gamma$ corresponding to Sample 1c originates from a larger fraction of β-phases after the recrystallizing process, which gives rise to a higher piezoelectricity. As for Sample 2 with larger diameter, $\gamma$ is 2.3 $\gamma_s$ for positive field and is 1.5 $\gamma_s$ for negative field. The slight difference between different field directions comes from two possible reasons: 1) when changing the direction of voltage, the remnant polarizability makes a compensate to the electric field poling effect; 2) when $\delta_1/\beta_1^2 \neq \delta_2/\beta_2^2$, $\gamma$ changes when the electric field change direction according to Eq. (5). It is interesting to point out that the thermal conductivity of Sample 1b does not change with electric field. This result is in consistent with the morphology of the sample, as it has been reported that P(VDF-TrFE) transforms from trans conformers (β-phase) to gauche isomers (α-phase) around 60°C in vacuum, and consequently the ferroelectricity will be largely reduced [32]. Macroscopically, Sample 1b has ignorable piezoelectric effect, thus $\gamma$ is too small to be observed. From above results, we find that the ferroelectricity is essential to achieve the phonon renormalization. Large crystallinity of ferroelectric-phase and better chain orientation are preferred.

In summary, we have revealed that the electric field could induce phonon renormalization in ferroelectric polymers. The measured thermal conductivity of ferroelectric P(VDF-TrFE) nanofibers exhibits a monotonic increasing behavior under electric field, which is in good agreement with our analytical phonon renormalization model. The tunable molecular chain structure under electric field and reversible phase transition is expected to be a promising prospect in the future application field of polymer ferroelectrics.

**Acknowledgements**
This work was supported by National Key R&D Program of China (No. 2017YFB0406004) and the National Natural Science Foundation of China (Grant No. 11890703, Grant No. 11674245)

**Figure Captures**

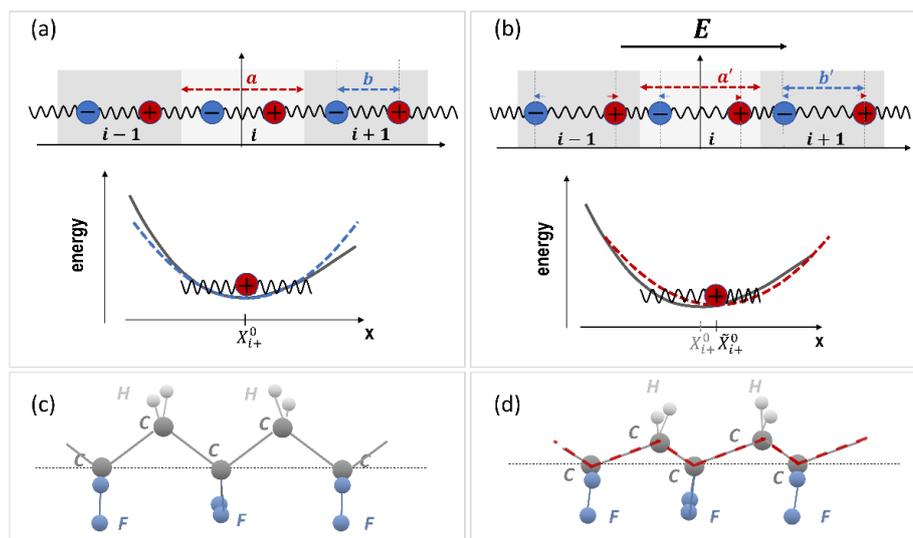

Fig.1 Schematic of the phonon renormalization in case of external electric field. (a) When there is no electric field, the diatomic chain is in equilibrium, with atoms locating at their equilibrium positions. The lattice constant is $a$ and the separation between two atoms in each unit cell is $b$. The potential energy of atoms with positive charge is plotted by the black solid line, and a harmonic approximation is plotted by the blue dash. (b) After applying the external electric field, the atoms in the diatomic chain will displace from the original equilibrium position. The lattice constant changes to $a'$ and the separation between two atoms in each unit cell becomes $b'$. The harmonic approximation of the potential at the new equilibrium position will be different from the original one due to the anharmonicity. (c) Typical molecular chain structure of PVDF or similar polymers. (d) In case of molecular chains, the variations of chain structure contain the change of bond angle and bond length.

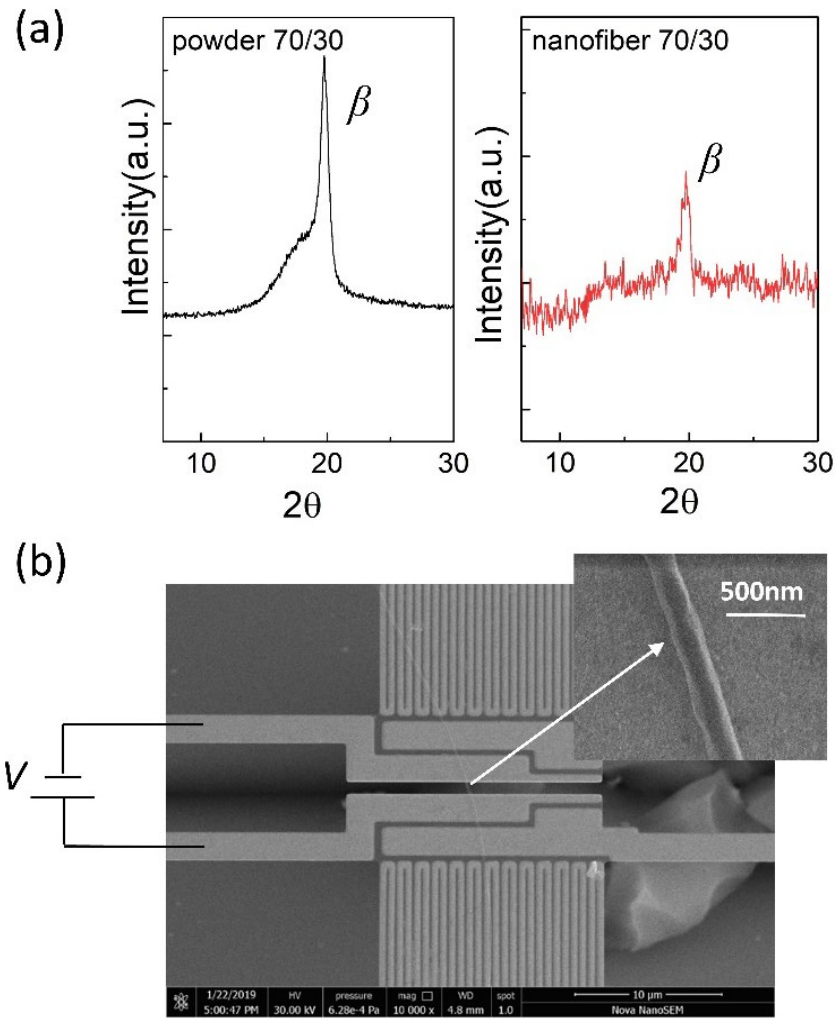

Fig.2 (a) X-ray diffraction results of P(VDF-TrFE) 70/30 powders and nanofibers. The marked β phase is generally considered as the ferroelectric phase. (b) SEM image of the single P(VDF-TrFE) nanofiber which suspended on the device for thermal conductivity measurement. The scale bar is 10 μm. Insert: the enlarged SEM image of the same P(VDF-TrFE) nanofiber. The scale bar is 500 nm.

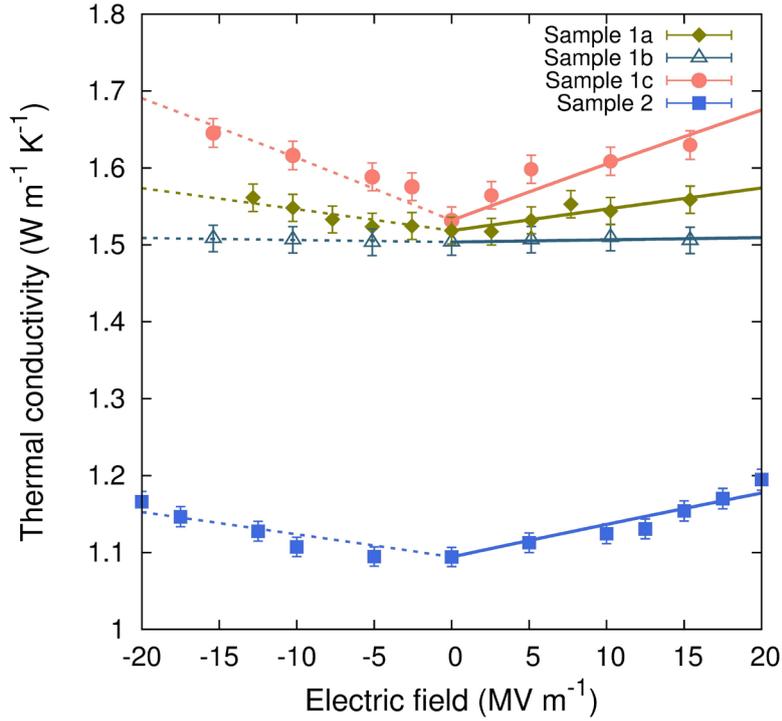

Fig.3 Electric filed dependence of the thermal conductivity at 300K of samples under different annealing conditions. Solid dots are measured data and lines are fitted by Eq. (8).

Table 1. The fitted $\gamma$ and $\gamma/\gamma_s$ of different P(VDF-TrFE) nanofibers.

| Sample | Annealing condition | $\kappa_0$ (Wm$^{-1}$K$^{-1}$) | $\gamma$ (nmV$^{-1}$) Positive field | $\gamma$ (nmV$^{-1}$) Negative field | $\gamma/\gamma_s$ Positive field | $\gamma/\gamma_s$ Negative field |
|---|---|---|---|---|---|---|
| 1a | 140°C; N$_2$; 30min furnace cooling | 1.52 | 3.7±0.5 | 3.7±0.4 | 1.1 | 1.1 |
| 1b | 67°C; vacuum; bake | 1.50 | 0.4±0.2 | 0.35±0.07 | - | - |
| 1c | 140°C; N$_2$; 2h slow cooling | 1.53 | 9.8±1.2 | 10.9±1.0 | 2.8 | 3.1 |
| 2 | 140°C; N$_2$; 30min furnace cooling | 1.10 | 7.9±0.6 | 5.5±0.7 | 2.3 | 1.5 |